\documentclass[12pt]{article}
\usepackage{amsfonts}
\usepackage{amssymb,amsmath,amsthm,latexsym}
\usepackage[numbers,compress]{natbib}

\newtheorem{theorem}{Theorem}[section]
\newtheorem{prop}[theorem]{Proposition}
\newtheorem{cor}[theorem]{Corollary}
\newtheorem{lemma}[theorem]{Lemma}

\newtheorem{remark}[theorem]{Remark}
\newtheorem{define}[theorem]{Definition}
\newtheorem{example}[theorem]{Example}

\newcommand{\EOP} { \hfill $\Box$ }

\newcommand{\pf} { {\rm \noindent{\bf Proof.}} }

\numberwithin{equation}{section}
\begin{document}
\title{   On More Bent Functions From Dillon   Exponents}
\author{Long Yu$^1$, Hongwei Liu$^{1}${\thanks{Corresponding author:~hwliu@mail.ccnu.edu.cn~}}~, Dabin Zheng$^2$ }
\date{${}^1$School of Mathematics and Statistics, Central China Normal University, Wuhan, Hubei 430079, China\\
 ${}^2$School of Mathematics and Statistics, Hubei University,  Wuhan, Hubei 430062, China}

\maketitle
\begin{abstract}
In this paper, we obtain a new class of $p$-ary binomial bent   functions which are determined by Kloosterman sums. The bentness of   another three  classes of functions is characterized  by some exponential sums and some results in \cite{Linian2013} are generalized. Furthermore we  obtain, in some special cases,  some  bent functions  are  determined by  Kloosterman sums.
\end{abstract}

%%%%%%%%%%%%%%%%%%%%%%%%%%%%%%%%%%%%%%%%%%%%%%%%%%%%%%%%%%%%%%%%%%%%%%%%%%

\section{Introduction}

In $1976$, Rothaus \cite{Rothaus1976} introduced boolean bent functions which are maximally nonlinear boolean functions with even number of variables. That is, they achieve the
maximal Hamming distance between boolean functions and affine functions. Boolean bent functions have attracted much attention due to their application in coding theory, cryptography and sequence design. Later, Kumar, Scholtz and Welch
 \cite{Kumar-Scholtz-Welch1985} generalized the notion of boolean bent functions to  the case of functions over an arbitrary finite field $\mathbb{F}_{p^n}$, where $p$ is a prime integer and $n$ is a positive integer. Some results on constructions of bent functions on monomial, binomial and quadratic functions  could be found in \cite{Canteaut2008,Charpin-Gong2008,Dobbertin-Leander-Canteaut-Carlet2006, Dillon1974, Helleseth-Kholosha2006,Helleseth-Kholosha2010, Jia-Zengetl2012,Leander2006,Linian2013,Mesnagerbent2011,Wang-Tang-Yang-Xu,zheng-yu-hu2013}.

Throughout this paper, let $p$ be a prime integer and $m$ be a positive integer with $n=2m$, $\mathbb{F}_{p^{n}}$ be the finite field with $p^n$ elements and $\mathbb{F}_{p^n}^*=\mathbb{F}_{p^n}\backslash \{0\}$. Let ${\rm Tr}_1^n(\cdot)$ be the trace function from $\mathbb{F}_{p^n}$ to $\mathbb{F}_p$, i.e. ${\rm Tr}_1^n(x)=\sum_{i=0}^{n-1}x^{p^i}$ for $x\in \mathbb{F}_{p^n}$.
The bentness of boolean monomials with Dillon exponents was characterized by Dillon in \cite{Dillon1974} and   Charpin et~al. in \cite{Charpin-Gong2008}.  The corrosponding $p$-ary case was investigated by Helleseth and Kholosha in \cite{Helleseth-Kholosha2006}. Some multinomial bent functions with Dillon exponents were investigated in \cite{Jia-Zengetl2012}, \cite{Mesnagerbent2011}, \cite{Wang-Tang-Yang-Xu}, \cite{zheng-yu-hu2013}. Recently, Li et~al. \cite{Linian2013} investigated the bentness of several special classes of functions in the following form
\begin{equation}\label{eq:fxzongdexingshi}
f(x)=\sum_{i=1}^{p^m-1}{\rm Tr}_1^n(a_i x^{i(p^m-1)})+{\rm Tr}_1^{o(\!d )}(b x^{\frac{p^n-1}{d}}),
\end{equation}
where $n=2m$, $a_i\in \mathbb{F}_{p^n},\, b \in \mathbb{F}_{p^{o(d)}} $,
$d$ is a positive integer with $d \mid (p^m+1)$ and $o(d)$ is the smallest positive integer
satisfying ${o(d)}\mid n$ and $d \mid (p^{o(d)}-1)$.   The bentness of all these special classes of  functions is determined by some exponential sums, most of which have close relations with Kloosterman sums.

The aim of this paper is that we further investigate four classes of bent functions in the form (\ref{eq:fxzongdexingshi}), which generalize some results in  \cite{Linian2013}. By applying the results on $S_i(a)$, $i=0,1$, $d=2$ (see \cite{Jia-Zengetl2012}),  we establish a relationship between  some partial exponential sums and Kloosterman sums. Based on this result, a new class of $p$-ary binomial bent functions are obtained (see Theorem~\ref{thm:pisoddsecondbent}).  Moreover,  we further investigate the bentness of another three classes of bent functions from Dillon exponents in the form (\ref{eq:fxzongdexingshi}). In particular,  the bentness of some   functions is determined by Kloosterman sums (see  Theorem~\ref{thm:pisoddsecondbent}, Corollaries~\ref{cor:1}, \ref{cor:2}, \ref{cor:3}).

% Using Lemma~\ref{lem:fbentzongkehua}, we can also characterized  several special classes of bent functions of (\ref{eq:fxzongdexingshi}) by some partial exponential sums. In particular, some bent functions can be   determined by the well-known Kloosterman sums().

The remainder of this paper is organized as follows. Section $2$ gives some preliminaries.  In Section $3$,  the bentness of four  classes of  functions is characterized by  some  exponential sums.  The concluding remarks are given in Section $4$.

\section{Preliminaries }
In this section, we give some basic definitions and results.
\begin{define}Let $f:\mathbb{F}_{p^{n}}\rightarrow \mathbb{F}_{p}$ be a $p$-ary function. The {\it Walsh transform} of $f$ is defined by
 \[W_f(\lambda) = \sum_{ x\in \mathbb{F}_{p^n} } \omega^{f(x)-{\rm Tr}_1^n(\lambda x)},\,\, \lambda \in \mathbb{F}_{p^{n}} , \]
 where $\omega=e^{\frac{2\pi\sqrt{-1}}{p}}$ is a complex primitive $p$-th root of unity.
\end{define}
\begin{define}\label{def:bent}
Let $f: \mathbb{F}_{p^{n}}\rightarrow
\mathbb{F}_{p}$ be a $p$-ary function. Then $f(x)$ is called  a bent function if $|W_f(\lambda)|^2= p^n$ for all $\lambda\in \mathbb{F}_{p^n}$. A $p$-ary bent function $f(x)$ is said to be {\it regular} if
 for all $\lambda\in \mathbb{F}_{p^n}$, $W_f(\lambda) = p^{\frac{n}{2}}\omega^{f^*(\lambda)}$ for some functions $f^*$ from $\mathbb{F}_{p^n}$ to $\mathbb{F}_p$.
%A $p$-ary bent function $f(x)$ is called {\it weakly regular} if there exists a complex $\mu$ with unit magnitude such that
%$W_f(\lambda)=p^{\frac{n}{2}}\mu \omega^{f^*(\lambda)}$ for all $\lambda\in \mathbb{F}_{p^n}$.
The function $f^*(x)$ is called the {\it dual} of $f(x)$.
\end{define}
\begin{remark}
In particular, for $p=2$,  a boolean bent function is always regular.
\end{remark}
%3)\ For even $n$, a boolean function $f \ :\mathbb{F}_{2^{n}}\rightarrow
%\mathbb{F}_{2}$ is said to be semibent if $W_f(\lambda)\in \{0,\pm2^{\frac{n+2}{2}}\}$ for all $\lambda\in
%\mathbb{F}_{2^n}$. For odd $n$, a boolean function $f \ :\mathbb{F}_{2^{n}}\rightarrow
%\mathbb{F}_{2}$ is said to be semibent if $W_f(\lambda)\in \{0,\pm2^{\frac{n+1}{2}}\}$ for all $\lambda\in
%\mathbb{F}_{2^n}$.
%---------------------------------------------------------------------------------------------
%t is well known that a boolean function $f(x)$ only exists for even $n$ and it's maximal degree is $\frac{n}{2}$. For prime $p>2$, we have the following result of the degree of a $p$-ary bent function $f(x)$.
%
%\begin{prop}{\rm\cite{X.D.Hou2004}}\label{prop:daishucishu}
%If $f(x)$ is a $p$-ary bent function from $\mathbb{F}_{p^n}$ to $\mathbb{F}_p$, then the degree $deg(f)$ of $f(x)$ satisfies $deg(f)\leq \frac{(p-1)n}{2}+1$. Furthermore, if $f(x)$ is a regular bent function, then $deg(f)\leq \frac{(p-1)n}{2}$.
%\end{prop}
\begin{define}
The  Dickson polynomial $D_r(x)\in \mathbb{F}_{2}[x]$ of degree $r$ is defined by
\[D_r(x)=\sum_{i=0}^{\lfloor r/2\rfloor}\frac{r}{r-i}\left(
                                                     \begin{array}{c}
                                                       r-i \\
                                                       i \\
                                                     \end{array}
                                                   \right)x^{r-2i},\, \, \, \,r=2,3,\cdots,
 \]where $\left(
                                                     \begin{array}{c}
                                                       k \\
                                                       s \\
                                                     \end{array}\right)=\frac{\prod_{j=0}^{s-1}(k-j)}{\prod_{j=1}^{s}j}$ and $\lfloor \frac{r}{2}\rfloor=\left\{
                                                                                       \begin{array}{ll}
                                                                                         r/2, & \hbox{if $r$ is even;} \\
                                                                                         (r-1)/2, & \hbox{otherwise.}
                                                                                       \end{array}
                                                                                     \right.
                                                     $
\end{define}

\begin{define}Let $\alpha\in \mathbb{F}_{p^m}$, the Kloosterman sum $K_m(\alpha)$ over $\mathbb{F}_{p^m}$ is defined as
$$K_m(\alpha)=\sum_{x\in \mathbb{F}_{p^m}}\omega^{Tr_1^m(\alpha x+x^{p^m-2})},$$
where $\omega=e^{\frac{2\pi\sqrt{-1}}{p}}$ is a complex primitive $p$-th root of unity.
\end{define}
\noindent It is easy to see that $K_m(\alpha)$ is a real number, where $\alpha\in \mathbb{F}_{p^m}$, $m$ is a positive integer.

%In order to  investigate the  bent functions given  in Section $3$, we  consider some exponential sums in the following.
% For convenience, we give some notations.
Let  $d$ be a  divisor of $p^m+1$ and
$U=\{ x\mid x^{p^m+1}=1, x\in \mathbb{F}_{p^n}\}$ be a cyclic subgroup of $\mathbb{F}_{p^n}^*$.
It is easy to check that $U$   can be decomposed into $U=\bigcup_{k=0}^{d-1}V_k,$
where $V_0=\{\xi^{di}\mid 0\leq i < \frac{p^m+1}{d}\}$, $V_k=\xi^k V_0$ for $ 1\leq k\leq d-1$, and $\xi$ is a generator of the cyclic group $U$.
For $i=0,1,\cdots,d-1$ and $a\in \mathbb{F}_{p^n}$, we define
\[S_i(a)=\sum_{x\in V_0}\omega^{{\rm Tr}_1^n(a\xi^ix)}.\]
It is well known that if $p=2$, then $$\sum_{x\in U}\omega^{{\rm Tr}_1^n(ax)}=\sum_{i=0}^{d-1}S_i(a)=1-K_m(a),\ \ \ a \in \mathbb{F}_{2^m}.$$
If  $p>2$, then $$\sum_{x\in U}\omega^{{\rm Tr}_1^n(ax)}=\sum_{i=0}^{d-1}S_i(a)=1-K_m(a^{p^m+1}),\ a\in \mathbb{F}_{p^n}.$$
%Actually, the bentness of $f(x)$ defined by (\ref{eq:fxzongdexingshi}) is  determined by (\ref{eq:s(a-1a-2...a-n)}) (see Lemma~\ref{lem:fbentzongkehua}).
%However, for a  general $d$, it is difficult to obtain the value of  $S_i(a)$.
which is given in \cite{Helleseth-Kholosha2006}. In particular, for the case of  $p=2$ and $d=3,$ Mesnager \cite{Mesnagersemibent2011} found a relationship between $S_i(a)$ and some well-known  exponential sums, and then constructed  a new class of binomial bent functions. Furthermore, for $p=2$ and $d=5$, $S_i(a)$ was  determined by some well-known  exponential sums in \cite{Wang-Tang-Yang-Xu}. Using these results, they also characterized the bentness of  a new class of binomial  functions.
For  $p>2$, the only known results on  $S_i(a)$  were given in \cite{Jia-Zengetl2012}, where $i=0,1$ and $d=2$, which were used to characterize a new class of binomial  bent functions. Following this idea, the bentness of more  functions  can be characterized if $S_i(a)$ is obtained for some $p$ and $d$, where $0\leq i\leq d-1$.

In particular, for $p=2$, Li  et al. in \cite{Linian2013}   obtained  a relation between  $S_0(a)$ and some exponential sums as follows.
\begin{lemma}{\rm \cite{Linian2013}}\label{lem:S0}
Let $p=2$, $d$ be a divisor of $p^m+1$, $a=\overline{a}\xi^k\in \mathbb{F}_{2^n}, \overline{a}\in \mathbb{F}_{2^m}^*,\, 0\leq k\leq 2^m$. If $k\equiv 0 \ ({\rm mod} \ d)$, then
\[S_0(a)=\frac{1+2E_{m,d}(\overline{a})-K_m(\overline{a})}{d},\]
where $E_{m,d}(\overline{a})=\sum\limits_{x\in \mathbb{F}_{2^m}}(-1)^{{\rm Tr}_1^m(\overline{a}D_d(x))}$, $\xi$ is a generator of the cyclic group $U$.
\end{lemma}

For convenience, we  give some  notations.
Let  $p$ be an odd prime and $\alpha$ be a primitive element in $\mathbb{F}_{p^n}$. We define $\mathcal{C}_t=\{\alpha^{2i+t}\mid i=0,1,\cdots,\frac{p^n-3}{2}\}\subseteq \mathbb{F}_{p^n}^*\,{\rm for}\, t=0,1$.   For $a\in \mathbb{F}_{p^n}$ and  for $b\in \mathcal{C}_0$, we define $R(a)$ and $Q(b)$ as follows:
$$R(a)=\frac{1-K_m(a^{p^m+1})}{2},\,\,\,\,\,\,Q(b)=2{\rm Tr}_1^m(b^{\frac{p^m+1}{2}}).$$

\begin{lemma}\label{lem:tisone}{\rm \cite{Jia-Zengetl2012}} With the notations given above. Then for $d=2$, we have
\[S_0(a) = \left\{
   \begin{array}{ll}
     R(a)+ I (\omega^{Q(a)}- \omega^{-Q(a)}), & \hbox{$a\in \mathcal{C}_0^+ $;} \\
     R(a), & \hbox{otherwise,}
   \end{array}
 \right.\]
and
\[S_1(a) = \left\{
   \begin{array}{ll}
     R(a)- I (\omega^{Q(a)}- \omega^{-Q(a)}), & \hbox{$a\in \mathcal{C}_0^+ $;} \\
     R(a), & \hbox{otherwise,}
   \end{array}
 \right.\]
where
$I=\left\{
  \begin{array}{ll}
    \frac{(-1)^{\frac{3m}{2}}p^{\frac{m}{2}}}{2}, & \hbox{$p \equiv 3 \ ({\rm mod}\ 4)$;} \\
    \frac{(-1)^{m}p^{\frac{m}{2}}}{2}, & \hbox{otherwise,}
  \end{array}
\right.$and $\,\,\mathcal{C}_0^+=\{a\in \mathcal{C}_0\mid  ~Q(a)\neq0\}.$
\end{lemma}

In particular, let $p^m \equiv 3 \ ({\rm mod}\ 4)$, $d=4$, then the following relationship between $S_i(a)$, $i=1,3$, and Kloosterman sum can be  established.
\begin{cor}\label{cor:S1S3}
Let $\alpha$ be a primitive element of $\mathbb{F}_{p^n}$, $p^m \equiv 3 \ ({\rm mod}\ 4)$, $d=4$  and $a=\alpha^{i(p^m+1)}\in \mathbb{F}_{p^m}^*$, where $i$ is a positive integer with $0 \leq i \leq p^m-2$. Then
\[S_1(a)=S_3(a)=
     \frac{R(a)- I (\omega^{Q(a)}- \omega^{-Q(a)})}{2},\]where $R(a)$, $Q(a)$, $I$ are given in Lemma~\ref{lem:tisone}.
\end{cor}
\pf Since $p^m \equiv 3 \ ({\rm mod}\ 4)$ , then we have $4\mid (p^m+1)$ and
\begin{eqnarray*}
  S_1(a) &=& \sum\limits_{x\in V_0}\omega^{{\rm Tr}_1^n(a\xi x)}
  =  \sum\limits_{x\in V_0}\omega^{{\rm Tr}_1^n(a\xi^{p^m} x^{p^m})}
  =  \sum\limits_{x\in V_0}\omega^{{\rm Tr}_1^n(a\xi^{3}\xi^{p^m-3} x)}
  = \sum\limits_{x\in V_0}\omega^{{\rm Tr}_1^n(a\xi^{3} x)}\\
   &=&S_3(a).
\end{eqnarray*}
Since $S_1(a)+S_3(a)=\sum\limits_{x\in H}\omega^{{\rm Tr}_1^n(a\xi x)}$, where $H=\{\xi^{2i}\mid 0\leq i\leq \frac{p^m-1}{2}\}$, by  Lemma~\ref{lem:tisone}, we have
$$S_1(a) = S_3(a)=\left\{
   \begin{array}{ll}
     \frac{R(a)- I (\omega^{Q(a)}- \omega^{-Q(a)})}{2}, & \hbox{$a\in \mathcal{C}_0^+ $;} \\
     \frac{R(a)}{2}, & \hbox{otherwise.}
   \end{array}
 \right.$$where $R(a)$, $Q(a)$, $I$ are given in Lemma~\ref{lem:tisone}.
Note that $a\in \mathcal{C}_0$, we have that $Q(a)\neq0$ if $a\in \mathcal{C}_0^+$, and  $Q(a)=0$ if $a\in \mathcal{C}_0$ and $ a\notin \mathcal{C}_0^+$. This  finishes the proof.\EOP

%In what follows, we present   a necessary and sufficient condition such that $f(x)$ of the form (\ref{eq:fxzongdexingshi}) is bent in \cite{Linian2013} and give another approach to prove it for prime $p>2$.

Let $\alpha$ be a primitive element of $\mathbb{F}_{p^n}$. For an odd prime $p$,   every $x\in \mathbb{F}_{p^n}^*$ has a unique representation as $x=uy$, where $u\in \mathcal{U}=\{1,\alpha,\cdots,\alpha^{p^m}\}$, $y\in \mathbb{F}_{p^m}^*$. Then we  get the following proposition.
\begin{prop}\label{prop:pisoddweiyijie}
For $\lambda\in \mathbb{F}_{p^n}^*$, then there exists only one solution in $\mathcal{U}$ such that ${\rm Tr}_m^n(\lambda x)=0$.
\end{prop}

The following lemma can be found in \cite{Helleseth-Kholosha2006}.
\begin{lemma}\label{lemma:pisoddfisbent}{\rm \cite{Helleseth-Kholosha2006}}
Let $p$ be an odd prime, $f:\ \mathbb{F}_{p^n} \ \rightarrow \ \mathbb{F}_p$
be a regular bent function such that $f(x)=f(-x)$ and $f(0)=0$,
then $f^*(0)=0$, where $f^*$ is the dual function
of $f$.
\end{lemma}
A necessary and sufficient condition such that $f(x)$ defined by (\ref{eq:fxzongdexingshi}) is bent was given in \cite{Linian2013}. We restate this result and give  another proof.
\begin{lemma}\label{lem:fbentzongkehua}{\rm \cite{Linian2013}}
Assume the notations given above. Then the function $f(x)$ defined by (\ref{eq:fxzongdexingshi}) is regular  if and only if $$S(a_1,a_2,\cdots,a_{p^m-1},b)=1,$$ where $
    S(a_1,\cdots,a_{p^m-1},b)=\sum\limits_{x\in U}\omega^{\sum_{i=1}^{p^m-1}{\rm Tr}_1^n(a_ix^i)+{\rm Tr}_1^{o(\!d)}(bx^{\frac{p^m+1}{d}})}.
$
\end{lemma}
\pf We first compute the walsh transform of $f(x)$. If $\lambda=0$, then
\begin{eqnarray}\label{eq:Wf0}
\nonumber% \nonumber to remove numbering (before each equation)
W_f(0) &=&\sum\limits_{x\in\mathbb{F}_{p^n}}\omega^{f(x)}\\\nonumber
       &=& 1+\sum\limits_{u\in \mathcal{U}}\sum\limits_{y\in \mathbb{F}_{p^m}^*}\omega^{\sum\limits_{i=1}^{p^m-1}{\rm Tr}_1^n(a_iu^{i(p^m-1)})+{\rm Tr}_1^{o(\!d)}(bu^{\frac{p^n-1}{d}})}  \\\nonumber
     &=& 1+(p^m-1)\sum\limits_{u\in \mathcal{U}}\omega^{\sum\limits_{i=1}^{p^m-1}{\rm Tr}_1^n(a_iu^{i(p^m-1)})
     +{\rm Tr}_1^{o(\!d)}(bu^{\frac{p^n-1}{d}})}\\\nonumber
     &=& 1+(p^m-1)\sum\limits_{x\in U}\omega^{\sum\limits_{i=1}^{p^m-1}{\rm Tr}_1^n(a_ix^{i})
     +{\rm Tr}_1^{o(\!d)}(bx^{\frac{p^m+1}{d}})}\\
     &=& 1+(p^m-1)S(a_1,a_2,\cdots,a_{p^m-1},b).\,
\end{eqnarray}
If $\lambda \in \mathbb{F}_{p^n}^*$, then
\begin{eqnarray}\label{eq:Wflamd}
W_f(\lambda) &=&\sum\limits_{x\in\mathbb{F}_{p^n}}\omega^{f(x)-{\rm Tr_1^n}(\lambda x)}\nonumber\\
       &=& 1+\sum\limits_{u\in \mathcal{U}}\sum\limits_{y\in \mathbb{F}_{p^m}^*}\omega^{\sum\limits_{i=1}^{p^m-1}{\rm Tr}_1^n(a_iu^{i(p^m-1)})+{\rm Tr}_1^{o(\!d)}(bu^{\frac{p^n-1}{d}})-{\rm Tr_1^n}(\lambda uy)}  \nonumber\\
     &=& 1+\sum\limits_{u\in \mathcal{U}}\sum\limits_{y\in \mathbb{F}_{p^m}}\omega^{\sum\limits_{i=1}^{p^m-1}{\rm Tr}_1^n(a_iu^{i(p^m-1)})
     +{\rm Tr}_1^{o(\!d)}(bu^{\frac{p^n-1}{d}})-{\rm Tr_1^n}(\lambda uy)}\nonumber\\
     & &-\sum\limits_{u\in \mathcal{U}}\omega^{\sum\limits_{i=1}^{p^m-1}{\rm Tr}_1^n(a_iu^{i(p^m-1)})
     +{\rm Tr}_1^{o(\!d)}(bu^{\frac{p^n-1}{d}})} \nonumber\\
     &=& 1+\sum\limits_{u\in \mathcal{U}}\omega^{\sum\limits_{i=1}^{p^m-1}{\rm Tr}_1^n(a_iu^{i(p^m-1)})
     +{\rm Tr}_1^{o(\!d)}(bu^{\frac{p^n-1}{d}})}\! \sum\limits_{y\in \mathbb{F}_{p^m}}\omega^{{-\rm Tr_1^m}(y{\rm Tr_m^n}(\lambda u))}\nonumber\\
     & &-\sum\limits_{x\in U}\omega^{\sum\limits_{i=1}^{p^m-1}{\rm Tr}_1^n(a_ix^{i})
     +{\rm Tr}_1^{o(\!d)}(bx^{\frac{p^m+1}{d}})}\nonumber\\
     %&=& 1+\sum\limits_{u\in \mathcal{U}}\omega^{\sum\limits_{i=1}^{p^m-1}{\rm Tr}_1^n(a_iu^{i(p^m-1)})
%     +{\rm Tr}_1^{o(\!d)}(bu^{\frac{p^n-1}{d}})}\!\sum\limits_{y\in \mathbb{F}_{p^m}}\omega^{{-\rm Tr_1^n}(\lambda uy)}\\
%     & &-S(a_1,a_2,\cdots,a_{p^m-1},b)\\
     &=& 1+p^m\omega^{\sum\limits_{i=1}^{p^m-1}{\rm Tr}_1^n(a_iu_{\lambda}^{i(p^m-1)})
     +{\rm Tr}_1^{o(\!d)}(bu_{\lambda}^{\frac{p^n-1}{d}})}-S(a_1,a_2,\cdots,a_{p^m-1},b)\,,
\end{eqnarray}
where $u_{\lambda}$ satisfies ${\rm Tr}_m^n(\lambda u)=0$ and the last equality in (\ref{eq:Wflamd}) is obtained by
Proposition~\ref{prop:pisoddweiyijie}.

Case I. $p=2$: if  $S(a_1,a_2,\cdots,a_{2^m-1},b)=1$,  it is easy to see that $f(x)$ is bent from equation~(\ref{eq:Wf0}) and (\ref{eq:Wflamd}). Conversely, if $f(x)$ is bent, then $W_f(0)=1+(2^m-1)S(a_1,a_2,\cdots,a_{2^m-1},b)$ $\in \{\pm 2^m\}$. Since $S(a_1,a_2,\cdots,a_{2^m-1},b)$ is an integer, then $S(a_1,a_2,\cdots,a_{2^m-1},b)=1$.

Case II. $p>2$: if  $f(x)$ is regular bent, then $W_f(0)=p^{m}\omega^{f^*(0)}$ by Definition~\ref{def:bent}. By Lemma~\ref{lemma:pisoddfisbent},
we have $W_f(0)=1+(p^m-1)S(a_1,a_2,\cdots,a_{p^m-1},b)$ $=p^m$.
Therefore, we  get $S(a_1,a_2,\cdots,a_{p^m-1},b)=1$.
Conversely, if $S(a_1,a_2,\cdots,a_{p^m-1},b)=1$, it is easy to check that $f(x)$ is  regular bent from equation~(\ref{eq:Wf0}) and (\ref{eq:Wflamd}).\EOP
%\begin{remark}
%For $p=2$, $f(x)$ is a regular bent function which is equivalent to $f(x)$ is a  bent function.
%\end{remark}
%------------------------------------------------------------------------------------------------------------------
\section{Binary and $p$-ary Bent Functions }
In this section, we study four classes of functions in the form  (\ref{eq:fxzongdexingshi}), whose bentness are determined by some  exponential sums.
%Moreover, some of results obtained in this section (see Remark~\ref{remark:1}, \ref{remark:2}, \ref{remark:3}, \ref{remark:4}), generalize the results of \cite{Linian2013}. In particular, we obtain more bent functions which are determined by the well-known Kloosterman sum.
\subsection{Binary bent functions}
In this subsection, we investigate   two special classes of bent functions in the form (\ref{eq:fxzongdexingshi}).

\subsubsection{First class of binary bent functions}
In the following of this part, we always assume that $d$ and  $l$  are  positive integers with $\gcd(l,\frac{2^m+1}{d})=1$.  We consider the bentness of the following functions
\begin{equation}\label{eq:firstclassbent}
f_{a_0,\cdots,a_{d-1},b}(x)=\sum_{i=0}^{d-1}{\rm Tr}_1^n(a_ix^{(l+i\frac{2^m+1}{d})(2^m-1)})+{\rm Tr}_1^{o(\!d)}(bx^{\frac{2^n-1}{d}}),
\end{equation}where $a_i\in \mathbb{F}_{2^n}$, $0\leq i\leq d-1$, $b\in \mathbb{F}_{2^{o(\!d)}}$.
\begin{theorem}\label{th:firstclasschongyao}
Assume the notations given above. Then the function $f_{a_0,\cdots,a_{d-1},b}(x)$  defined by (\ref{eq:firstclassbent})  is  bent if and only if
\[\sum\limits_{j=0}^{d-1}(-1)^{{\rm Tr}_1^{o(\!d)}(b\xi^{\frac{j(2^m+1)}{d}})}\sum\limits_{x\in V_0}(-1)^{\sum\limits_{i=0}^{d-1}{\rm Tr}_1^n(a_i\xi^{j(i\frac{2^m+1}{d})}\xi^{jl} x)}=1.\]
\end{theorem}
\pf Note that
\begin{eqnarray}\label{eq:Sa0a1a2...ad-1}
\nonumber& & \sum\limits_{x\in U}(-1)^{\sum\limits_{i=0}^{d-1}{\rm Tr}_1^n(a_ix^{l+i\frac{2^m+1}{d}})+{\rm Tr}_1^{o(\!d)}(bx^{\frac{2^m+1}{d}})} \\\nonumber
&=& \sum\limits_{x\in V_0}(-1)^{\sum\limits_{i=0}^{d-1}{\rm Tr}_1^n(a_ix^l)+{\rm Tr}_1^{o(\!d)}(b)}+\sum\limits_{x\in V_0}(-1)^{\sum\limits_{i=0}^{d-1}{\rm Tr}_1^n(a_i\xi^{l+i\frac{2^m+1}{d}}x^l)+{\rm Tr}_1^{o(\!d)}(b\xi^{\frac{2^m+1}{d}})} \\\nonumber
& &+\cdots+\sum\limits_{x\in V_0}(-1)^{\sum\limits_{i=0}^{d-1}{\rm Tr}_1^n(a_i\xi^{(d-1)(l+i\frac{2^m+1}{d})}x^l)+{\rm Tr}_1^{o(\!d)}(b\xi^{\frac{(d-1)(2^m+1)}{d}})} \\\nonumber
&=& \sum\limits_{x\in V_0}(-1)^{\sum\limits_{i=0}^{d-1}{\rm Tr}_1^n(a_ix)+{\rm Tr}_1^{o(\!d)}(b)}+\sum\limits_{x\in V_0}(-1)^{\sum\limits_{i=0}^{d-1}{\rm Tr}_1^n(a_i\xi^{i\frac{2^m+1}{d}}\xi^l x)+{\rm Tr}_1^{o(\!d)}(b\xi^{\frac{2^m+1}{d}})}\\\nonumber
& &+\cdots+\sum\limits_{x\in V_0}(-1)^{\sum\limits_{i=0}^{d-1}{\rm Tr}_1^n(a_i\xi^{(d-1)(i\frac{2^m+1}{d})}\xi^{(d-1)l} x)+{\rm Tr}_1^{o(\!d)}(b\xi^{\frac{(d-1)(2^m+1)}{d}})}\\
&=&\sum\limits_{j=0}^{d-1}(-1)^{{\rm Tr}_1^{o(\!d)}(b\xi^{\frac{j(2^m+1)}{d}})}\sum\limits_{x\in V_0}(-1)^{\sum\limits_{i=0}^{d-1}{\rm Tr}_1^n(a_i\xi^{j(i\frac{2^m+1}{d})}\xi^{jl} x)}.
\end{eqnarray}
By Lemma~\ref{lem:fbentzongkehua}, we  finish the proof.\EOP

%Proposition~\ref{prop:firstclasschongyao} is a description of bentness of $f(x)$ defined by (\ref{eq:firstclassbent}).
Moreover, for some case of $a_i$'s,   the bentness of $f_{a_0,\cdots,a_{d-1},b}(x)$ defined by (\ref{eq:firstclassbent}) can be characterized by some well-known  exponential sums.
\begin{theorem}\label{thm:diyileixishu}
Let $a_0\in\mathbb{F}_{2^m}^*,\,a_1=a_2=\cdots=a_{d-1}\in\mathbb{F}_{2^m}$ and $a_0\neq a_1$. Then $f_{a_0,\cdots,a_{d-1},0}(x)$ be defined by (\ref{eq:firstclassbent}) is bent if  and only if
\begin{small}$$K_m(a_0)+(d-1)K_m(a_0+a_1)=\left\{
  \begin{array}{ll}
    2(E_{m,d}(a_0)+(d-1)E_{m,d}(a_0+a_1)), & \hbox{{\rm if} $d\mid l$;} \\
    2(E_{m,d}(a_0)-E_{m,d}(a_0+a_1)), & \hbox{{\rm if} $\gcd(d,l)=1$,}
  \end{array}
\right.$$\end{small}where $E_{m,d}(a)$ is given in Lemma~\ref{lem:S0}.
\end{theorem}
\pf Since  $\xi^{j\frac{2^m+1}{d}}$ is a root of $1+z+z^2+\cdots+z^{d-1}=0$ for each $1\leq j \leq d-1$, and $a_1=a_2=\cdots=a_{d-1}$, we get  $\sum\limits_{i=0}^{d-1}{\rm Tr}_1^n(a_i\xi^{i(j\frac{2^m+1}{d})}\xi^{jl} x)={\rm Tr}_1^n\left((a_0+a_1)\xi^{jl}x\right)$ for each $1\leq j \leq d-1$. Note that $b=0$ and $\gcd(l,\frac{2^m+1}{d})=1$, then Equation (\ref{eq:Sa0a1a2...ad-1}) is
\begin{eqnarray}\label{eq:a0=a1=anS}
\nonumber   & &\sum\limits_{x\in U}(-1)^{\sum\limits_{i=0}^{d-1}{\rm Tr}_1^n(a_ix^{l+i\frac{2^m+1}{d}})}\\
       %&=& \sum\limits_{z\in U}(-1)^{\sum\limits_{i=0}^{d-1}{\rm Tr}_1^n(a_iz^{l+i\frac{2^m+1}{d}})} \\ \\
%     &=& \sum\limits_{z\in V_0}(-1)^{{\rm Tr}_1^n(a_0z^l+ \sum\limits_{i=1}^{d-1}a_1z^{l})}+\sum\limits_{z\in V_0}(-1)^{{\rm Tr}_1^n(a_0\xi^lz^l
%+ \sum\limits_{i=1}^{d-1}a_1\xi^{l+i\frac{2^m+1}{d}}z^l)}\\ \\
%     & & +\cdots+ \sum\limits_{z\in V_0}(-1)^{{\rm Tr}_1^n(a_0\xi^{(d-1)l}z^l+ \sum\limits_{i=1}^{d-1}a_1\xi^{(d-1)(l+i\frac{2^m+1}{d})}z^l)}\\ \\
%&=&\sum\limits_{z\in V_0}(-1)^{{\rm Tr}_1^n((a_0+ \sum\limits_{i=1}^{d-1}a_1)z^{l})}+ \sum\limits_{z\in V_0}(-1)^{{\rm Tr}_1^n((a_0+ \sum\limits_{i=1}^{d-1}a_1\xi^{i\frac{2^m+1}{d}})\xi^lz^{l})}\\ \\
%& & +\cdots+ \sum\limits_{z\in V_0}(-1)^{{\rm Tr}_1^n((a_0+ \sum\limits_{i=1}^{d-1}a_1\xi^{i\frac{2^m+1}{d}(d-1)})\xi^{(d-1)l}z^{l})} \\ \\
&=& \sum\limits_{x\in V_0}(-1)^{{\rm Tr}_1^n(a_0x)}+\sum\limits_{x\in V_0}(-1)^{{\rm Tr}_1^n((a_0+a_1)\xi^lx)}+\cdots+\sum\limits_{x\in V_0}(-1)^{{\rm Tr}_1^n((a_0+a_1)\xi^{(d-1)l}x)}.
\end{eqnarray}

In the following, we discuss Equation (\ref{eq:a0=a1=anS}) in two cases.

1) If $d\mid l$, then by Lemma~\ref{lem:S0}, we have
\begin{eqnarray*}
  & &\sum\limits_{x\in U}(-1)^{\sum\limits_{i=0}^{d-1}{\rm Tr}_1^n(a_ix^{l+i\frac{2^m+1}{d}})}\\
&=&\sum\limits_{x\in V_0}(-1)^{{\rm Tr}_1^n(a_0x)}+(d-1)\sum\limits_{x\in V_0}(-1)^{{\rm Tr}_1^n((a_0+a_1)x)}\\
&=& \frac{1+2E_{m,d}(a_0)-K_m(a_0)}{d}+(d-1)\frac{1+2E_{m,d}(a_0+a_1)-K_m(a_0+a_1)}{d}.
\end{eqnarray*}

Hence, by Theorem~\ref{th:firstclasschongyao}, $f_{a_0,\cdots,a_{d-1},b}(x)$ is bent if and only if
\[K_m(a_0)+(d-1)K_m(a_0+a_1)=2(E_{m,d}(a_0)+(d-1)E_{m,d}(a_0+a_1)).\]

2) If $\gcd(d,l)=1 $, it is easy to verify that $\{l \ ({\rm mod}\ d),\,2l\ ({\rm mod}\ d),\,\cdots,\,(d-1)l\ ({\rm mod}\ d)\}=\{1,\,2,\,\cdots,\,d-1\}$. By Lemma~\ref{lem:S0},  we have
\begin{eqnarray*}
 & &\sum\limits_{x\in U}(-1)^{\sum\limits_{i=0}^{d-1}{\rm Tr}_1^n(a_ix^{l+i\frac{2^m+1}{d}})}\\
&=&\sum\limits_{x\in V_0}(-1)^{{\rm Tr}_1^n(a_0x)}+\sum\limits_{x\in V_0}(-1)^{{\rm Tr}_1^n((a_0+a_1)\xi x)}+\cdots+\sum\limits_{x\in V_0}(-1)^{{\rm Tr}_1^n((a_0+a_1)\xi^{d-1} x)}\\
                    &=&\sum\limits_{x\in V_0}(-1)^{{\rm Tr}_1^n(a_0x)}+\sum\limits_{x\in V_1}(-1)^{{\rm Tr}_1^n((a_0+a_1)x)} +\cdots+\sum\limits_{x\in V_{d-1}}(-1)^{{\rm Tr}_1^n((a_0+a_1) x)}\\
                    &=&\sum\limits_{x\in V_0}(-1)^{{\rm Tr}_1^n(a_0x)}+\sum\limits_{x\in U}(-1)^{{\rm Tr}_1^n((a_0+a_1)x)}-\sum\limits_{x\in V_0}(-1)^{{\rm Tr}_1^n((a_0+a_1)x)}\\
                    &=&1-K_m(a_0+a_1)+\frac{1+2E_{m,d}(a_0)-K_m(a_0)}{d}-\frac{1+2E_{m,d}(a_0+a_1)-K_m(a_0+a_1)}{d}.
\end{eqnarray*}
Therefore, by Theorem~\ref{th:firstclasschongyao}, $f_{a_0,\cdots,a_{d-1},b}(x)$ is bent if and only if
\[K_m(a_0)+(d-1)K_m(a_0+a_1)=2E_{m,d}(a_0)-2E_{m,d}(a_0+a_1).\]
This finishes the proof.\EOP

If we take $d=3$ in Theorem~\ref{thm:diyileixishu}, and combine the results on $S_i(a)$ in \cite{Mesnagersemibent2011}, $i=0,1,2$, we  obtain the following result, which is exactly  Corollary~1 in \cite{Linian2013}.
\begin{cor}\label{cor:remark1}
Let $f_{a_0,\cdots,a_{d-1},0}(x)$ be defined by (\ref{eq:firstclassbent}) with $b=0$, $d=3$, $a_0\in\mathbb{F}_{2^m}^*,\,a_1=a_2\in\mathbb{F}_{2^m}$ and $a_0\neq a_1$. Then $f_{a_0,\cdots,a_{d-1},0}(x)$ is bent if  and only if
\begin{small}$$K_m(a_0)+2K_m(a_0+a_1)=\left\{
  \begin{array}{ll}
    2\left(C_{m}(a_0)+2C_{m}(a_0+a_1)\right), & \hbox{{\rm if} $3\mid l$;} \\
    2\left(C_{m}(a_0)-C_{m}(a_0+a_1)\right), & \hbox{{\rm  otherwise},}
  \end{array}
\right.$$\end{small}where $C_{m}(a)=\sum_{a\in \mathbb{F}_{2^m}}(-1)^{{\rm Tr}_1^m(ax^3+ax)}$.
\end{cor}

\begin{example}
Let $n=2m=6$, $d=9$ and $l=1$, $a_0\in \mathbb{F}_{2^3}^*$, $a_1=a_2=\cdots=a_8\in \mathbb{F}_{2^3}^* $, then $f_{a_0,\cdots,a_{8},0}(x)={\rm Tr}_1^6(a_0x^{7})+\sum_{i=1}^{8}{\rm Tr}_1^6(a_1x^{7(1+i)})$. By using Maple, we get that there  exist $9$ pairs $(a_0,a_1)$ such that $f_{a_0,\cdots,a_{8},0}(x)$ is bent.
\end{example}

If $b\neq0$, by a similar discussion as that in Theorem~\ref{thm:diyileixishu}, we  obtain the following theorem.
\begin{theorem}\label{thm:diyileixishuyoub}
Let $f_{a_0,\cdots,a_{d-1},b}(x)$ be defined by (\ref{eq:firstclassbent}) with $b\neq0$,  $d\mid l$, $a_0\in\mathbb{F}_{2^m}^*,\,a_1=a_2=\cdots=a_{d-1}\in\mathbb{F}_{2^m}$ and $a_0\neq a_1$. Then $f_{a_0,\cdots,a_{d-1},b}(x)$ is bent if  and only if $$\rho K_m(a_0)+\sigma K_m(a_0+a_1)= 2(\rho E_{m,d}(a_0)+\sigma E_{m,d}(a_0+a_1))+\rho+\sigma-d,$$where $\rho=(-1)^{{\rm Tr}_1^{o(\!d)}(b)}$, $\sigma=\sum\limits_{j=1}^{d-1}(-1)^{{\rm Tr}_1^{o(\!d)(b\xi^{j\frac{2^m+1}{d}})}}$ and $E_{m,d}(a)$ is given in Lemma~\ref{lem:S0}.
\end{theorem}
If we take $a_1=a_2=\cdots=a_{d-1}=0$, then we have  the following result by Theorem~\ref{thm:diyileixishuyoub}. which is exactly Theorem~3 in \cite{Linian2013}.
\begin{cor}\label{cor:remark2}
Let $d \mid l$, $a_0\in\mathbb{F}_{2^m}^*$ and $a_1=\cdots=a_{d-1}=0$. Then $f_{a_0,0,\cdots,0,b}(x)$ defined by (\ref{eq:firstclassbent}) is bent if and only if
\[\sum_{j=0}^{d-1}(-1)^{{\rm Tr}_1^{o(\!d)}(b\xi^{j\frac{2^m+1}{d}})}=\frac{d}{1+2E_{m,d}(a_0)-K_m(a_0)}.\]
where $E_{m,d}(a)$ is given in Lemma~\ref{lem:S0}.\end{cor}

\begin{example}
Let $n=2m=4$, $d=5$ and $l=5$, $\alpha$ be a primitive element of $\mathbb{F}_{2^4}$, $a_0\in \mathbb{F}_{2^2}^*$, $a_0\neq a_1$, $a_1=a_2=a_3=a_{4}\in \mathbb{F}_{2^2}^* $, $b\in  \mathbb{F}_{2^4}^*$, then $f_{a_0,\cdots,a_4,b}(x)$ defined by (\ref{eq:firstclassbent}) is equal to
${\rm Tr}_1^4(a_0x^{15})+\sum_{i=1}^{4}{\rm Tr}_1^4(a_1x^{3(5+i)})+{\rm Tr}_1^4(bx^{3})$. By using Maple, the number of $(a_0,a_1,b)$ such that $f_{a_0,\cdots,a_4,b}(x)$ is bent function is $60$.
\end{example}
\subsubsection{Second class of binary bent functions }
In this part, we always assume that $s$, $k$, $r$  are integers with $r\mid (2^m+1)$.
We investigate  the bentness of
\begin{equation}\label{eq:dierlei}
f_{a,r,s}(x)=\sum_{i=1}^{\frac{2^m+1}{r}-1}{\rm Tr}_1^n(ax^{(ri+s)(2^m-1)}),%{\rm Tr}_1^{o(\!d)}(bx^{t\frac{2^n-1}{d}}),
\end{equation}
where $a\in\mathbb{F}_{2^n}^*$ and $f(0)=0$.

%Using Lemma~\ref{lem:fbentzongkehua}, the bentness of $f_{a,r,s}(x)$ defined by (\ref{eq:dierlei}) can be characterized by some exponential sums.
 %As we know, the terms of $f(x)$ defined by (\ref{eq:dierlei}) may produce constant when $ri+s=k(2^m+1)$, $k$ is an integer. Moreover,  $f(x)$ is bent if and only if $f(x)+c$ is bent, where $c \in \mathbb{F}_2$ is a constant. Since  $f(x)$
%in Lemma~\ref{lem:fbentzongkehua} dose not contain constant terms,   thus we  minus the constant terms of $f(x)$, i.e. $f(x)-f(0)$, and investigate the bentness of $f(x)-f(0)$.
 %which is equivalent to the bentness of $f(x)$.
\begin{theorem}\label{thm:dierleihanshu}
Assume the notations given above. Then
\begin{enumerate}
  \item if $\gcd(s,2^m+1)=1$, $0\leq k\leq 2^m$ and $a=\overline{a}\xi^k\in\mathbb{F}_{2^n}$ with $\overline{a}\in \mathbb{F}_{2^m}^*$, then $f_{a,r,s}(x)$ is  bent if and only if
$$K_m(\overline{a})=r-\sum\limits_{x^r=1,x\in U}(-1)^{{\rm Tr}_1^n(ax)}.$$
  \item if $\gcd(s,2^m+1)=d$, $0\leq k< \frac{2^m+1}{d}$ and $a=\overline{a}\xi^{kd} \in\mathbb{F}_{2^n}$ with $\overline{a}\in \mathbb{F}_{2^m}^*$, then $f_{a,r,s}(x)$ is bent if and only if
$$dS_0(\overline{a})=\sum\limits_{x^r=1,x\in U}(-1)^{{\rm Tr}_1^n(ax^s)}+1-r,$$where $S_0(\overline{a})$ is given in Lemma~\ref{lem:S0}.
\end{enumerate}
\end{theorem}
\pf By Lemma~\ref{lem:fbentzongkehua}, $f_{a,r,s}(x)$ is bent if and only if
$\sum\limits_{x\in U}(-1)^{\sum\limits_{i=1}^{\frac{2^m+1}{r}-1}{\rm Tr}_1^n(ax^{ri+s})}=1.$
On the other hand, $\sum\limits_{i=1}^{\frac{2^m+1}{r}-1}{\rm Tr}_1^n(ax^{(ri+s)})={\rm Tr}_1^n(ax^s)$  when $x^r \neq 1$  and $x\in U$.
Since $\frac{2^m+1}{r}-1$ is even, we get
\begin{eqnarray*}
 \sum\limits_{x\in U}(-1)^{\sum\limits_{i=1}^{\frac{2^m+1}{r}-1}{\rm Tr}_1^n(ax^{ri+s})}&=&\sum\limits_{x\in U\setminus{x^r=1}}(-1)^{{\rm Tr}_1^n(ax^s)}+r\\
    &=&\sum\limits_{x\in U}(-1)^{{\rm Tr}_1^n(ax^s)}+r-\sum\limits_{x^r=1,x\in U}(-1)^{{\rm Tr}_1^n(ax^s)}.
\end{eqnarray*}

If $\gcd(s,2^m+1)=1$, then $\sum\limits_{x\in U}(-1)^{{\rm Tr}_1^n(ax^s)}=\sum\limits_{x\in U}(-1)^{{\rm Tr}_1^n(ax)}=1-K_m(\overline{a})$
 and $\gcd(s,r)=1$. Thus  $f_{a,r,s}(x)$ is bent if and only if $$K_m(\overline{a})=r-\sum\limits_{x^r=1,x\in U}(-1)^{{\rm Tr}_1^n(ax)}.$$

If $\gcd(s,2^m+1)=d$ and $a=\overline{a}\xi^{kd}$, we have $$\sum\limits_{x\in U}(-1)^{{\rm Tr}_1^n(ax^s)}=d\sum\limits_{x\in V_0}(-1)^{{\rm Tr}_1^n(ax)}=dS_0(\overline{a}) .$$
Thus $f_{a,r,s}(x)$ is bent if and only if $$dS_0(\overline{a})=\sum\limits_{x^r=1,x\in U}(-1)^{{\rm Tr}_1^n(ax^s)}+1-r. $$\EOP

In particular, if we take $r=1$ in Theorem~\ref{thm:dierleihanshu}, we have the following result, which is exactly Theorem~4 in \cite{Linian2013}.
\begin{cor}\label{cor:remark3}
Assume the notation given above. Then we have
\begin{enumerate}
  \item if $\gcd(s,2^m+1)=1$, $0\leq k\leq 2^m$ and $a=\overline{a}\xi^k\in\mathbb{F}_{2^n}$ with $\overline{a}\in \mathbb{F}_{2^m}$, then $f_{a,1,s}(x)$ is  bent if and only if
$$K_m(\overline{a})=1-(-1)^{{\rm Tr}_1^n(a)}.$$
  \item if $\gcd(s,2^m+1)=d$, $0\leq k< \frac{2^m+1}{d}$ and $a=\overline{a}\xi^{kd} \in\mathbb{F}_{2^n}$ with $\overline{a}\in \mathbb{F}_{2^m}$, then $f_{a,1,s}(x)$ is bent if and only if
$$dS_0(\overline{a})=(-1)^{{\rm Tr}_1^n(a)},$$where $S_0(\overline{a})$ is given in Lemma~\ref{lem:S0}.
\end{enumerate}
\end{cor}

In particular, let $r=3$, $\gcd(s,2^m+1)=1$, then we have the following corollary.
\begin{cor}\label{cor:1}
Assume the notation given above. Let $\gcd(s,2^m+1)=1$, $a=\overline{a}\xi^k\in\mathbb{F}_{2^n}$ with $\overline{a}\in \mathbb{F}_{2^m}$ and $0 \leq k\leq 2^m$, $f(0)=0$ and $r=3$. Then $f_{a,3,s}(x)$ defined by (\ref{eq:dierlei}) is  bent if and only if
$$K_m(\overline{a})=3-\sum\limits_{j=0}^2(-1)^{{\rm Tr}_1^n(a\xi^{j\frac{2^m+1}{3}})}.$$
Furthermore, if $f_{a,3,s}(x)$ is bent, then $K_m(\overline{a})=0$ when ${\rm Tr}_1^n(a)={\rm Tr}_1^n(a\xi^{\frac{2^m+1}{3}})={\rm Tr}_1^n(a\xi^{2\frac{2^m+1}{3}})=0$,  otherwise, $K_m(\overline{a})=4$.
\end{cor}
\pf By Theorem~\ref{thm:dierleihanshu}, we have that $f_{a,3,s}(x)$ defined by (\ref{eq:dierlei}) is  bent if and only if
$$K_m(\overline{a})=3-\sum\limits_{j=0}^2(-1)^{{\rm Tr}_1^n(a\xi^{j\frac{2^m+1}{3}})}.$$
Note that ${\rm Tr}_1^n(a)={\rm Tr}_1^n(a\xi^{\frac{2^m+1}{3}})+{\rm Tr}_1^n(a\xi^{2\frac{2^m+1}{3}})$, since $\xi^{\frac{2^m+1}{3}}+\xi^{2\frac{2^m+1}{3}}=1$. It is easy to check that $K_m(\overline{a})=0$ if ${\rm Tr}_1^n(a)={\rm Tr}_1^n(a\xi^{\frac{2^m+1}{3}})={\rm Tr}_1^n(a\xi^{2\frac{2^m+1}{3}})=0$ and in other cases, $K_m(\overline{a})=4$. This completes the proof. \EOP

\begin{example}
Let $\alpha$ be a primitive element of $\mathbb{F}_{2^6}$, $n=2m=6$, $r=3$, $s=1$, $a\in \mathbb{F}_{2^6}^*$, then $f_{a,3,1}(x)={\rm Tr}_1^6(ax^{28})+{\rm Tr}_1^6(ax^{49})$. By applying Maple, the number of   this class of binomial regular bent functions is $36$.
\end{example}

\subsection{p-ary bent functions}
In this subsection, we always assume that $p$ is an odd prime.
%Two special classes of $p$-ary bent functions in the form (\ref{eq:fxzongdexingshi})
%are investigated.
By  Corollary~\ref{cor:S1S3}, the bentness of a new class of binomial $p$-ary  functions (see  Theorem~\ref{thm:pisoddsecondbent}) is characterized by Kloosterman sums. Following the similar idea of construction of the second class of binary bent functions,
we also obtain another   class of regular bent functions (see  Theorem~\ref{thm:pisodddisanlei}), and in some special cases,
we  get  more bent functions which are determined by Kloosterman sums.
\subsubsection{First class of $p$-ary bent functions}
We have established a connection between  $S_i(a)$, $i=1,3$, and Kloosterman sum in Corollary~\ref{cor:S1S3} and  use this result to characterize  the bentness of the following binomial functions
\begin{equation}\label{eq:poddsecondbent}
f_{a,b}(x)={\rm Tr}_1^n(a x^{l(p^m-1)})+{\rm Tr}_1^2(b x^{\frac{p^n-1}{4}}),
\end{equation}
where $p^m\equiv3\ ({\rm mod}\ 4)$, $a\in \mathbb{F}_{p^n}^*$, $b\in \mathbb{F}_{p^2}^*$, $l$ is an integer with $\gcd(l,\frac{p^m+1}{4})=1$.

%If $b=0$, then $f(x)={\rm Tr}_1^n(a x^{l(p^m-1)})$, which has been investigated in \cite{Helleseth-Kholosha2006}. Therefore, we investigate the bentness of $f(x)$ defined by (\ref{eq:poddsecondbent}) for $b\neq0$.

%Thus, by Lemma~\ref{lem:fbentzongkehua}, the bentness of  $f(x)$ defined by (\ref{eq:poddsecondbent}) can be  characterized.
\begin{theorem}\label{pro:pisoddsecondbent}
Assume the notations given above. Then $f_{a,b}(x)$ defined by (\ref{eq:poddsecondbent}) is regular bent if and only if
\[\sum\limits_{j=0}^3\omega^{{\rm Tr}_1^2(b\xi^{j\frac{p^m+1}{4}})}\sum\limits_{x\in V_0}\omega^{{\rm Tr}_1^n(a\xi^{jl} x)}=1.\]
\end{theorem}
\pf Since
\begin{eqnarray*}
& & \sum\limits_{x\in U}\omega^{{\rm Tr}_1^n(a x^{l})+{\rm Tr}_1^2(b x^{\frac{p^m+1}{4}})}\\
               \vspace{2mm} &=&\sum\limits_{x\in V_0}\omega^{{\rm Tr}_1^n(a x^{l})+{\rm Tr}_1^2(b)}+\sum\limits_{x\in V_0}\omega^{{\rm Tr}_1^n(a \xi^lx^{l})+{\rm Tr}_1^2(b\xi^{\frac{p^m+1}{4}})}\\
& &+\sum\limits_{x\in V_0}\omega^{{\rm Tr}_1^n(a\xi^{2l}x^{l})-{\rm Tr}_1^2(b)}+\sum\limits_{x\in V_0}\omega^{{\rm Tr}_1^n(a \xi^{3l} x^{l})+{\rm Tr}_1^2(b\xi^{3\frac{p^m+1}{4}})} \\
               \vspace{2mm} &=& \omega^{{\rm Tr}_1^2(b)}\sum\limits_{x\in V_0}\omega^{{\rm Tr}_1^n(a x)}+\omega^{{\rm Tr}_1^2(b\xi^{\frac{p^m+1}{4}})}\sum\limits_{x\in V_0}\omega^{{\rm Tr}_1^n(a \xi^lx)}\\
& &+\omega^{{\rm Tr}_1^2(-b)}\sum\limits_{x\in V_0}\omega^{{\rm Tr}_1^n(a \xi^{2l}x)}+\omega^{{\rm Tr}_1^2(b\xi^{3\frac{p^m+1}{4}})}\sum\limits_{x\in V_0}\omega^{{\rm Tr}_1^n(a\xi^{3l} x)}\\
&=& \sum\limits_{j=0}^3\omega^{{\rm Tr}_1^2(b\xi^{j\frac{p^m+1}{4}})}\sum\limits_{x\in V_0}\omega^{{\rm Tr}_1^n(a\xi^{jl} x)}.
         \end{eqnarray*}
Then, by Lemma~\ref{lem:fbentzongkehua}, we have that $f_{a,b}(x)$ is regular bent if and only if $$\sum\limits_{j=0}^3\omega^{{\rm Tr}_1^2(b\xi^{j\frac{p^m+1}{4}})}\sum\limits_{x\in V_0}\omega^{{\rm Tr}_1^n(a\xi^{jl} x)}=1.$$\EOP

In particular,   we obtain the following result.
\begin{theorem}\label{thm:pisoddsecondbent}
Assume the notations given above. Let $k$ be a positive integer with $k \equiv 1 \,{\rm or}\, 3\ ({\rm mod}\ 4)$, $a=\overline{a}\xi^k$, where $\overline{a}\in \mathbb{F}_{p^m}^*$, and $4 \mid l $. Then $f_{a,b}(x)$ defined by (\ref{eq:poddsecondbent}) is regular bent if and only if
\[K_m(\overline{a}^2)=
     1-4I\sqrt{-1}\sin{\frac{2\pi Q(\overline{a})}{p}}-\frac{2}{\cos{\frac{2 \pi {\rm Tr}_1^2(b)}{p}}+\cos{\frac{2 \pi {\rm Tr}_1^2(b\xi^{\frac{p^m+1}{4}})}{p}}} \ ,\]where $Q(\overline{a})$, $I$ are given in Lemma~\ref{lem:tisone}.
\end{theorem}
\pf By  Theorem~\ref{pro:pisoddsecondbent}, we have that $f_{a,b}(x)$ is regular bent if and only if
\[\sum\limits_{j=0}^3\omega^{{\rm Tr}_1^2(b\xi^{j\frac{p^m+1}{4}})}\sum\limits_{x\in V_0}\omega^{{\rm Tr}_1^n(a\xi^{jl} x)}=1.\]
On the other hand, since  $4 \mid l $, then  \[\sum\limits_{j=0}^3\omega^{{\rm Tr}_1^2(b\xi^{j\frac{p^m+1}{4}})}\sum\limits_{x\in V_0}\omega^{{\rm Tr}_1^n(a\xi^{jl} x)}=\sum\limits_{j=0}^3\omega^{{\rm Tr}_1^2(b\xi^{j\frac{p^m+1}{4}})}\sum\limits_{x\in V_0}\omega^{{\rm Tr}_1^n(ax)}.\]
Since $a=\overline{a}\xi^k$, $ \overline{a} \in \mathbb{F}_{p^m}^*$, $k \equiv 1 \,{\rm or}\, 3\ ({\rm mod}\ 4)$, we have that
\begin{equation*}\label{eq:S1=S3}
\sum\limits_{x\in V_0}\omega^{{\rm Tr}_1^n(ax)}=\sum\limits_{x\in V_0}\omega^{{\rm Tr}_1^n(\overline{a} \xi x)}=\sum\limits_{x\in V_0}\omega^{{\rm Tr}_1^n(\overline{a} \xi^3 x)}=S_1(\overline{a})=S_3(\overline{a}).
\end{equation*}
Thus, by Corollary~\ref{cor:S1S3},
we have
\begin{equation*}\label{eq:xindeqiuhe}
\sum\limits_{x\in V_0}\omega^{{\rm Tr}_1^n(ax)}=
     \frac{R(\overline{a})- I (\omega^{Q(\overline{a})}- \omega^{-Q(\overline{a})})}{2}.
\end{equation*}
To sum up, $f_{a,b}(x)$ is regular bent if and only if
\[\sum\limits_{j=0}^3\omega^{{\rm Tr}_1^2(b\xi^{j\frac{p^m+1}{4}})}=
     \frac{4}{1-K_m(\overline{a}^2)-4I\sqrt{-1}\sin{\frac{2\pi Q(\overline{a})}{p}}},\]where $Q(\overline{a})$, $I$ are given in Lemma~\ref{lem:tisone}.
Note that $$\sum\limits_{j=0}^3\omega^{{\rm Tr}_1^2(b\xi^{j\frac{p^m+1}{4}})}=2({\cos{\frac{2 \pi {\rm Tr}_1^2(b)}{p}}
+\cos{\frac{2 \pi {\rm Tr}_1^2(b\xi^{\frac{p^m+1}{4}})}{p}}}),$$ we finish the proof.\EOP

\begin{cor}
If there exist $(a,b)\in\mathbb{F}_{3^n}^*\times \mathbb{F}_{3^2}^*$ such that $f_{a,b}(x)$ defined by (\ref{eq:poddsecondbent}) is a regular bent function. Then the number of these regular bent functions is divided by $4$.
\end{cor}
\pf Since $b\in \mathbb{F}_{3^2}^*$, then we can get $b\in \{\alpha^{i\frac{3^n-1}{8}}\mid 0\leq i\leq 7\}$, where $\alpha$ is a primitive element in $\mathbb{F}_{3^n}$. Since $\xi$ is the generator of $U$, so $\xi^{\frac{3^m+1}{4}}=\alpha^{\frac{3^n-1}{4}}$. Then $b$, $b\alpha^{\frac{3^n-1}{4}}$, $-b$, $b\alpha^{3\frac{3^n-1}{4}}$ have the same value of ${\cos{\frac{2 \pi {\rm Tr}_1^2(b)}{3}}
+\cos{\frac{2 \pi {\rm Tr}_1^2(b\xi^{\frac{p^m+1}{4}})}{3}}}$. This completes the proof.\EOP

%%%%%%%%%%%%%%%%%%%%%%%%%%%%%%%%%%%%%%%%%%%%%%%%%%%%%%%%%%%%%%%%%%%%%
\begin{example}
Let  $l=4$, $a=\overline{a}\xi$, $\overline{a}\in \mathbb{F}_{3^3}^* $, $b\in \mathbb{F}_{3^2}^*$, $\xi$ be a generator of cyclic  $U=\{x\in \mathbb{F}_{3^6} \mid x^{3^3+1}=1\}$, then we have $3^3 \equiv 3 \ ({\rm mod}\ 4)$ and $f_{a,b}(x)={\rm Tr}_1^6(ax^{144})+{\rm Tr}_1^2(bx^{182})$. By using Maple, the number of this binomial bent functions is $48$.
\end{example}
\begin{remark}
Following the similar construction of bent functions in Theorem~\ref{thm:diyileixishu} and Theorem~\ref{thm:diyileixishuyoub}, we can also investigate the bentness of  $f(x)=\sum_{i=0}^3{\rm Tr}_1^n(a_i x^{(l+i\frac{p^m+1}{4})(p^m-1)})+{\rm Tr}_1^2(b x^{\frac{p^n-1}{4}})$. In particular, by  Corollary~\ref{cor:S1S3}, the bentness of this class of functions can also be characterized by some exponential sums, which have close relations with Kloosterman sums.
\end{remark}
\subsubsection{Second  class of $p$-ary bent functions}
In this part, we always assume  $s$, $r$ are integers with $\gcd(s,p^m+1)=1$ and  $r \mid (p^m+1)$. Similar to the second class of binary bent functions, we investigate the bentness of the following function
\begin{equation}\label{eq:pisodddthirduoxiang}
    f_{a,b,r}(x)=\sum_{i=1}^{\frac{p^m+1}{r}-1}{\rm Tr}_1^n(ax^{(ri+s)(p^m-1)})+bx^\frac{p^n-1}{2},
\end{equation}
where $a\in \mathbb{F}_{p^n}^*$, $b\in \mathbb{F}_p^*$    and $f(0)=0$.\\
\begin{theorem}\label{thm:pisodddisanlei}
Assume the notations given above. Then $f_{a,b,r}(x)$ defined by (\ref{eq:pisodddthirduoxiang}) is regular bent  if and only if
\[(1-K_m(a^{p^m+1}))\cos\frac{2\pi b}{p}=\left\{
  \begin{array}{ll}
    4I\sin\frac{2\pi b}{p}\sin\frac{2\pi Q(-a)}{p}+\epsilon, & \hbox{$-a\in \mathcal{C}_0^+$;} \\
    \epsilon, & \hbox{otherwise,}
  \end{array}
\right.\]
where $\epsilon=\sum\limits_{x^r=1,x\in U}\omega^{{\rm Tr}_1^n(-ax^s)+bx^{\frac{p^m+1}{2}}}-\sum\limits_{x^r=1,x\in U}\omega^{(\frac{p^m+1}{r}-1){\rm Tr}_1^n(ax^s)+bx^{\frac{p^m+1}{2}}}+1.$
\end{theorem}
\pf Note that if $ x^r\neq1 $ and $ x\in U $, then
$
\sum\limits_{i=1}^{\frac{p^m+1}{r}-1}{\rm Tr}_1^n(x^{(ri+s)})
=-{\rm Tr}_1^n(x^s).
$
By Lemma~\ref{lem:fbentzongkehua}, $f_{a,b,r}(x)$ is regular bent if and only if
\[\sum_{x\in U}\omega^{\sum\limits_{i=1}^{\frac{p^m+1}{r}-1}{\rm Tr}_1^n(ax^{(ri+s)})+bx^{\frac{p^m+1}{2}}}=1,\]
which is equivalent to

\begin{equation}\label{eq:poddthirdbbudengyu0}
\sum_{x^r=1,x\in U}\omega^{(\frac{p^m+1}{r}-1){\rm Tr}_1^n(ax^s)+bx^{\frac{p^m+1}{2}}}+\sum_{x^r\neq1,x\in U}\omega^{{\rm Tr}_1^n(-ax^s)+bx^{\frac{p^m+1}{2}}}=1.
\end{equation}
On the other hand, we have
\begin{eqnarray}\label{eq:poddthirdbbudengyu1}
& &\sum\limits_{x^r\neq1,x\in U}\omega^{{\rm Tr}_1^n(-ax^s)+bx^{\frac{p^m+1}{2}}}\nonumber \\
&=&\sum\limits_{x\in U}\omega^{{\rm Tr}_1^n(-ax^s)+bx^{\frac{p^m+1}{2}}}-\sum\limits_{x\in U,x^r=1}\omega^{{\rm Tr}_1^n(-ax^s)+bx^{\frac{p^m+1}{2}}}\nonumber \\
&=&\!\omega^b\!\sum\limits_{x\in V_0}\!\omega^{{\rm Tr}_1^n(-ax^s)}\!+\!\omega^{-b}\!\sum\limits_{x\in V_0}\omega^{{\rm Tr}_1^n(-a\xi^sx^s)}\!-\!\sum\limits_{x^r=1,x\in U}\!\omega^{{\rm Tr}_1^n(-ax^s)+bx^{\frac{p^m+1}{2}}}\!.
\end{eqnarray}
Since $\gcd(s,p^m+1)=1$, then
\begin{eqnarray}\label{eq:poddthirdbbudengyu2}
   & & \omega^b\sum\limits_{x\in V_0}\omega^{{\rm Tr}_1^n(-ax^s)}+\omega^{-b}\sum\limits_{x\in V_0}\omega^{{\rm Tr}_1^n(-a\xi^sx^s)} \nonumber \\
   &=&  \omega^b\sum\limits_{x\in V_0}\omega^{{\rm Tr}_1^n(-ax)}+\omega^{-b}\sum\limits_{x\in V_0}\omega^{{\rm Tr}_1^n(-a\xi^sx)}\nonumber \\
&=&  \omega^b\sum\limits_{x\in V_0}\omega^{{\rm Tr}_1^n(-ax)}+\omega^{-b}\sum\limits_{x\in V_0}\omega^{{\rm Tr}_1^n(-a\xi(\xi^{s-1}x))}\nonumber \\
&=& \omega^bS_0(-a)+\omega^{-b}S_1(-a).
\end{eqnarray}
From (\ref{eq:poddthirdbbudengyu0}), (\ref{eq:poddthirdbbudengyu1}), (\ref{eq:poddthirdbbudengyu2}) and  Lemma~\ref{lem:tisone}, we complete this proof.\EOP

In particular, we have the following result, which is exactly Theorem~$10$, $11$ in \cite{Linian2013}.
\begin{cor}\label{cor:remark4}
Assume the notations given above. We have
\begin{enumerate}
  \item if $b=0$, $r=1$, then $f_{a,0,1}(x)$ is regular bent if and only if $K_m(a^{p^m+1})=1-\omega^{{\rm Tr}_1^n(-a)}.$

  \item if $b\neq0$, $r=1$, then $f_{a,b,1}(x)$ is regular bent if and only if
  $$(1-K_m(a^{p^m+1}))\cos\frac{2\pi b}{p}=\left\{
  \begin{array}{ll}
    4I\sin\frac{2\pi b}{p}\sin\frac{2\pi Q(-a)}{p}+\epsilon, & \hbox{$-a\in \mathcal{C}_0^+$;} \\
    \epsilon, & \hbox{otherwise,}
  \end{array}
\right. $$where $\epsilon=\omega^{{\rm Tr}_1^n(-a)+b}-\omega^{b}+1$.
\end{enumerate}
\end{cor}

Note that if $p=3$, $r=2$, then $\frac{p^m+1}{r}-1=\frac{3^m+1}{2}-1=1+3+\cdots+3^{m-1} \equiv1\ ( {\rm mod} \ 3)$. Together with $\gcd(s,3^m+1)=1$ and $b=0$, we have
\[\begin{array}{rcl}
\epsilon &=&\sum\limits_{x^2=1,x\in U}\omega^{{\rm Tr}_1^n(-ax^s)}-\sum\limits_{x^2=1,x\in U}\omega^{(\frac{3^m+1}{2}-1){\rm Tr}_1^n(ax^s)}+1\\
&=& \sum\limits_{x=\pm1}\omega^{{\rm Tr}_1^n(-ax^s)}-\sum\limits_{x=\pm1}\omega^{{\rm Tr}_1^n(ax^s)}+1\\
&=&\sum\limits_{x=\pm1}\omega^{{\rm Tr}_1^n(-ax)}-\sum\limits_{x=\pm1}\omega^{{\rm Tr}_1^n(ax)}+1\\
&=&1.
\end{array}\]

Therefore, we have the following result immediately.
\begin{cor}\label{cor:2}
Assume the notations given above. Let $p=3$,  $r=2$ and $b=0$. Then $f_{a,0,2}(x)$ defined by (\ref{eq:pisodddthirduoxiang}) is regular bent if and only if
\[K_m(a^{3^m+1})=0.\]
\end{cor}
%\begin{remark}
%As we known,for $a\in \mathbb{F}_{3^n}$,there are exist many solutions of $K_m(a^{3^m+1})$ $=0$. Thus the above bent functions can be obtained by the relations of $K_m(a^{3^m+1})=0$, which implies that  the above regular bent function exists.
%\end{remark}

Note that if $3^m \equiv3 \ ({\rm mod}\ 4)$, one has $\frac{3^m+1}{2}-1=1+3+\cdots+3^{m-1} \equiv1\ ({\rm mod}\ 3)$ and $\frac{3^m+1}{2}$ is an even integer. Together with $\gcd(s,3^m+1)=1$, we have
\[\begin{small}\begin{array}{rcl}
\epsilon &=&\sum\limits_{x^2=1,x\in U}\omega^{{\rm Tr}_1^n(-ax^s)+bx^{\frac{3^m+1}{2}}}-\sum\limits_{x^2=1,x\in U}\omega^{(\frac{3^m+1}{2}-1){\rm Tr}_1^n(ax^s)+bx^{\frac{3^m+1}{2}}}+1\\
&=& \sum\limits_{x=\pm1}\omega^{{\rm Tr}_1^n(-ax^s)+bx^{\frac{3^m+1}{2}}}-\sum\limits_{x=\pm1}\omega^{{\rm Tr}_1^n(ax^s)+bx^{\frac{3^m+1}{2}}}+1\\
&=&\omega^b\sum\limits_{x=\pm1}\omega^{{\rm Tr}_1^n(-ax)}-\omega^b\sum\limits_{x=\pm1}\omega^{{\rm Tr}_1^n(ax)}+1\\
&=&1.
\end{array}\end{small}\]

Therefore, we  have the following result.
\begin{cor}\label{cor:3}
Assume the notations given above. Let $b\neq0$, $p=3$,  $3^m \equiv3 \mod (4)$ and $r=2$. Then $f_{a,b,2}(x)$ defined by (\ref{eq:pisodddthirduoxiang}) is regular bent if and only if $$K_m(a^{p^m+1})=1-\frac{1}{\cos\frac{2\pi b}{p}}.$$
\end{cor}
\pf Since $\epsilon=1$, by  Theorem~\ref{thm:pisodddisanlei}, we have that $f_{a,b,2}(x)$ is regular bent if and only if
\[(1-K_m(a^{p^m+1}))\cos\frac{2\pi b}{p}=\left\{
  \begin{array}{ll}
    4I\sin\frac{2\pi b}{p}\sin\frac{2\pi Q(-a)}{p}+1, & \hbox{$-a\in \mathcal{C}_0^+$;} \\
    1, & \hbox{otherwise.}
  \end{array}
\right.\]
Since $p=3$, $3^m \equiv3 \mod (4)$, then $I$ is a complex number in Lemma~\ref{lem:tisone} and note that $(1-K_m(a^{p^m+1}))\cos\frac{2\pi b}{p}$ is a real number. Thus  $$(1-K_m(a^{p^m+1}))\cos\frac{2\pi b}{p}=-4I\sqrt{-1}\sin\frac{2\pi b}{p}\sin\frac{2\pi Q(-a)}{p}+1$$ if and only if $Q(-a)=0$, which contradicts with $-a\in \mathcal{C}_0^+$. That is to say $f_{a,b,2}(x)$ can not be bent if $-a\in \mathcal{C}_0^+$. This finishes the proof.\EOP

%\begin{remark}
%%Let $\alpha$ be a primitive element of $\mathbb{F}_{3^6}$, $r=2$, $s=1$ and then $f(x)=\sum_{i=1}^{13}{\rm Tr}_1^6(ax^{26(2i+1)})+bx^{314}$. There exists $(a,b)$ such that $f(x)$ is regular bent.
%The key proof of Corollary~3 is that if $f(x)$ is regular bent, then $a \notin \mathcal{C}_0^+$, which has been detailed proved in \cite{zheng-yu-hu2013}. Moreover, the previous works(please see {\rm \cite{Jia-Zengetl2012},\cite{zheng-yu-hu2013}})  have given that there exists  solutions of $(1-K_m(a^{p^m+1}))\cos\frac{2\pi b}{p}=1 $ for $p=3$, which implies there exists this kind of regular bent functions in Corollary~$3$. Furthermore,  the number of this kind of regular bent functions is even.
%\end{remark}

%%%%%%%%%%%%%%%%%%%%%%%%%%%%%%%%%%%%%%%%%%%%%%%%%%

\section{Concluding Remarks}
In this paper, several new  classes of  binary and $p$-ary bent functions with Dillon exponents are obtained. The bentness of all these functions are
characterized by some  exponential sums. Moreover, some of results obtained in this paper  generalize the work of  \cite{Linian2013}.

%%%%%%%%%%%%%%%%%%%%%%%%%%%%%%%%%%%%%%%%


\begin{thebibliography}{99}
%\bibitem{Carlet-Ding2004} C. Carlet and C. Ding,  Highly nonlinear mappings,  J. Complexity, vol. 20, no. 2, pp. 205-244, 2004.
\bibitem{Canteaut2008} A. Canteaut, P. Charpin and G. Kyureghyan, ``A new class of monomial bent functions'',
\emph{Finite Fields Appl}., \textbf{14}(1), 221-241, 2008.
\bibitem{Charpin-Gong2008} P. Charpin and G. Gong, ``Hyperbent functions, Kloosterman sums and
Dickson polynomials'', {\it IEEE Trans. Inf. Theory}, {\bf 9}(54), 4230-4238,  2008.

\bibitem{Dobbertin-Leander-Canteaut-Carlet2006} H. Dobbertin, G. Leander, A. Canteaut, C. Carlet, P. Felke and P. Gaborit, ``Construction of bent functions via Niho power functions'', {\it J.
Combin. Theory Ser. A,}   {\bf 113}(5), 779-798, 2006.

\bibitem{Dillon1974} J.  Dillon, ``Elementary Hadamard Difference Sets'', Ph.D. dissertation, Univ. Maryland, College Park, 1974.

\bibitem{Helleseth-Kholosha2006} T. Helleseth and A. Kholosha, ``Monomial and quadratic bent functions over the finite fields of odd characteristic'', {\it
 IEEE Trans. Inf. Theory,}   {\bf 52}(5),  2018-2032, 2006.

\bibitem{Helleseth-Kholosha2010}T. Helleseth and A. Kholosha, ``New binomial bent functions over finite fields of odd
characteristic'', \emph{IEEE Trans. Inf. Theory}, \textbf{56}(9), 4646-4652, 2010.

%\bibitem{Helleseth-Kholosha-Mesnager2012} T. Helleseth, A. Kholosha and S. Mesnager, Niho bent functions
%and Subiaco hyperovals, in proceedings of the 10-th International
%Conference on Finite Fields and Their Applications, Contemporary
%Math., AMS, 2012, to appear.




%\bibitem{X.D.Hou2004} X. Hou, p-ary and q-ary versions of certain resuils about bent functions and resilient functions, Finite Fields Appl.,  vol. 10, no. 4, pp. 566-582, 2004.

\bibitem{Jia-Zengetl2012} W.  Jia, X.  Zeng, T. Helleseth, and C.  Li, ``A class of binomial
bent functions over the finite fields of odd characteristic'', {\it IEEE Trans.
Inf. Theory,}  {\bf 58}(9), 6054-6063,  2012.

\bibitem{Kumar-Scholtz-Welch1985} P. V. Kumar, R. A. Scholtz, and L. R. Welch, ``Generalized bent functions and their properties'', {\it J. Combin. Theory Ser. A,}  {\bf 40}(1),   90-107, 1985.

\bibitem{Leander2006} G. Leander, ``Monomial bent functions'', \emph{IEEE Trans. Inform. Theory}, \textbf{52}(2), 738-743, 2006.

\bibitem{Linian2013} N. Li, T. Helleseth, X. Tang and A. Kholosha, ``Several new classes of bent functions from Dillon exponents'', {\it IEEE Trans. Inf. Theory,}  {\bf 59}(3), 1818-1831, 2013.



\bibitem{Mesnagersemibent2011} S. Mesnager, ``Semibent functions from Dillon and Niho exponents,
Kloosterman sums and Dickson polynomials'', {\it IEEE Trans. Inf. Theory,}  {\bf 57}(11), 7443-7458,  2011.

\bibitem{Mesnagerbent2011} S. Mesnager, ``Bent and hyper-bent functions in polynomial form and
their link with some exponential sums and Dickson polynomials'',
{\it IEEE Trans. Inf. Theory,}  {\bf 57}(9), 5996-6009,  2011.

%\bibitem{Rosendahl2006} P. Rosendahl, A generalization of Nihos theorem, Designs, Codes and
%Cryptography, vol. 38, no. 3, pp. 331-336, 2006.

\bibitem{Rothaus1976} O. S. Rothaus, ``On bent functions'', {\it J. Combin. Theory Ser. A,} {\bf 20}(3), 300-305, 1976.

\bibitem{Wang-Tang-Yang-Xu} C. Tang, Y. Qi, M. Xu, B. Wang and  Y. Yang,   ``A new class of hyper-bent Boolean functions in binomial forms [Online].'' Available: http://arxiv.org/pdf/1112.0062.pdf.

\bibitem{zheng-yu-hu2013} D. Zheng, L. Yu and L. Hu,
 ``On a class of binomial bent functions over the finite fields of odd characteristic'', {\it Applicable Algebra in Engineering, Communication and Computing,} {\bf 24}(6),  461-475, 2013.




\end{thebibliography}
\end{document}